\documentclass[journal]{IEEEtran}

\usepackage{cite}
\usepackage{array}
\usepackage{color}
\usepackage{float}
\usepackage[cmex10]{amsmath}
\usepackage{nomencl}						%nomenclature
\usepackage[normalem]{ulem}			%underline
\usepackage{diagbox}						%table head
\usepackage{slashbox}						%table head
\usepackage{colortbl}						%table color
\usepackage{multirow}						%table combine
\usepackage{tabularx}
\usepackage{placeins}
\usepackage{algorithm}
\usepackage[noend]{algpseudocode}
\usepackage{siunitx}
\usepackage{gensymb}

\ifCLASSINFOpdf
  \usepackage[pdftex]{graphicx}
	\graphicspath{{./figure/}}
  \DeclareGraphicsExtensions{.pdf,.jpeg,.png}
\else
  \usepackage[dvips]{graphicx}
  \graphicspath{{./figure/}}
  \DeclareGraphicsExtensions{.eps}
\fi

\ifCLASSOPTIONcompsoc
  \usepackage[caption=false,font=normalsize,labelfont=sf,textfont=sf]{subfig}
\else
  \usepackage[caption=false,font=footnotesize]{subfig}
\fi

\newcommand{\figref}[1]{\textcolor{black}{Fig.~\ref{#1}}}

\newcommand{\equref}[1]{\textcolor{black}{(\ref{#1})}}
\newcommand{\citeref}[1]{\textcolor{black}{\cite{#1}}}
\newcommand{\sectionref}[1]{\textcolor{black}{Section~\ref{#1}}}
\newcommand{\appendixref}[1]{\textcolor{black}{Appendix~\ref{#1}}}

\begin{document}
\bstctlcite{IEEEexample:BSTcontrol}
\setcounter{page}{1}

\title{Descriptor State Space Modeling of Power Systems}
\author{Yitong Li, \IEEEmembership{Member, IEEE}, Timothy C. Green, \IEEEmembership{Fellow, IEEE}, Yunjie Gu, \IEEEmembership{Senior Member, IEEE}}
%\thanks{Yitong Li, XXX and XXX are with the Department of Electrical and Electronic Engineering, Imperial College, London. E-mail: yitong.li15@imperial.ac.uk; yunjie.gu@imperial.ac.uk; t.green@imperial.ac.uk.}
%\thanks{This work was supported by...}

\ifCLASSOPTIONpeerreview
	\maketitle %\IEEEpeerreviewmaketitle
\else
	\maketitle
\fi

% ====================================================
% abstract
% ====================================================

\begin{abstract}
State space is widely used for modeling power systems and analyzing their dynamics but it is limited to representing causal and proper systems in which the number of zeros does not exceed the number of poles. In other words, the system input, output, and state can not be freely selected. This limits how flexibly models are constructed, and in some circumstances, can introduce errors because of the addition of virtual elements in order to connect the mismatched ports of subsystem models. An extension known as descriptor state space can model both proper and improper systems and is a promising candidate for solving the noted problems. It facilitates a modular construction of power system models with flexible choice of ports of subsystems. Algorithms for mathematical manipulation of descriptor state space models are derived such as preforming inverse, connection, and transform. Corresponding physical interpretations are also given. Importantly, the proposed algorithms preserve the subsystem states in the whole system model, which therefore enables the analysis of root causes of instability and mode participation. Theoretical advances are validated by example power systems of varied scales including a single-inductor or -capacitor system, two-inductor or -capacitor system, and a modified IEEE 14-bus generator-inverter-composite system.
\end{abstract}

% ====================================================
% Keywords
% ====================================================


% =========================================
% Introduction
% =========================================

\section{Introduction} \label{Section:Introduction}

For reliable provision of electricity services, power systems must remain stable and be capable of withstanding a wide range of disturbances \citeref{kundur1994power}. In addition to the traditional stability problems associated with synchronous generators (SGs), the increasing penetration of inverter-based resources (IBRs) lead to more stability threats \citeref{hatziargyriou2020definition}, partly because of the increases the system order (number of system states), and  participation analysis (root cause of oscillation modes) becomes more difficult. 

State space, as the most fundamental modeling method, has successfully been used for SG-based grids \citeref{kundur1994power}, and has more recently also been applied to IBR-dominated grids \citeref{markovic2019understanding,wang2017small,wang2018harmonic,pogaku2007modeling}. States in a state space model are usually linked to physical variables with clear practical interpretation, in other words it is a white-box model, and root causes of instability and oscillations modes can be identified based on modal and participation analysis \citeref{kundur1994power}. This can further guide parameter tuning and stability enhancement in practice. 

However, a normally overlooked fact is that the state space method can only model a \textit{proper} system in which the number of system zeros does not exceed the number of system poles \citeref{oppenheim1997signals,skogestad2005multivariable}. For example, when modeling the dynamics of an inductor in the Laplace $s$-domain, its current is the state and therefore has to be the system output, and its voltage has to be the system input. In other words, the system input, output, and state can not be freely selected. An important observation is that the subsystems in a practical power systems are normally ``terminal-inductive'' (e.g., filter inductance of IBRs, stator inductance of SGs, winding inductance of transformers, and line inductance of transmission and distribution networks). This means that the terminal voltage is always the input of a subsystem model and the terminal current is always the state and the output. This can lead to port mismatch when connecting subsystem state space models to construct the whole model of a multi-apparatus power system. In order to solve this problem, in \citeref{pogaku2007modeling,wang2017small}, a sufficiently-large virtual resistor is added at each bus node, to make the node voltage well-defined. Alternatively, in \citeref{markovic2019understanding,PSCAD}, each transmission and distribution line is assumed to be a $\pi$-section $CLC$ circuit, which is equivalent to adding a sufficiently-small virtual capacitor at each bus node. Obviously, the additional virtual resistors or capacitors change the physical structure of the system, introduce errors (maybe sufficiently-small but non-zero), and limit the flexibility with which models are constructed.

An alternative to state space modelling is the transfer function method which has successfully been used for analyzing dynamics of grid-connected power converters (normally using impedance or admittance transfer functions in electrical-centric view)\cite{wen2015analysis,wang2017unified,li2021impedance,fan2020identifying,liEArevisiting}. This approach has gained increasing interest recently because the electrical impedance spectrum can be directly measured in practice which enables the black-box analysis when physics-based models are not available \citeref{fan2020identifying,kontis2017measurement,li2022mapping}. The transfer function can naturally represent both proper and \textit{improper} systems.  However, when numerically modeling a high-order, multi-input, multi-output system, the transfer function normally has poorer numerical accuracy and is slower to compute than the equivalent state space model \citeref{MatlabModelConversion} because of: (a) symbolic calculation with Laplace operator $s$ for transfer function versus linear algebraic calculation with constant system matrices for state space; and (b) redundant system states and orders for different input-output combinations of transfer function versus shared state vector $x$ and state matrix $A$ for all inputs and outputs of state space. Additionally, the system state information is hidden in transfer function model due to the one-to-many mapping from it to a state space model, which complicates state participation analysis \citeref{kontis2017measurement,zhu2022participation,zhuEAimpedance}.

In this article, \textit{descriptor state space} (also known as \textit{implicit state space}) is introduced for modeling power systems. It is a generalized state space representation that can model both proper and improper systems. It was originally proposed many years ago in automatic control theory \citeref{luenberger1977dynamic,muller2000descriptor} but has only had very little attention in power engineering community, perhaps because electrical systems (actually all physically-realizable systems) can always be represented by proper state space models because they are casual systems in practice \citeref{oppenheim1997signals}. For example, a practical inductor can only be connected to a voltage source but not a current source, which implies an admittance model of a proper system $1/(sL)$ (voltage input and current output), and the corresponding state space model can not be inverted. As will be illustrated in this article, descriptor state space can push these limits which is in fact useful when modeling power systems because it provides a natural solution to the noted problems of port mismatch between subsystems with incompatible input-output definitions. To facilitate the formation of large power system models in an automated process, algorithms are created to manipulate descriptor state space subsystem models (such as inverse, transform, connection, etc) and physical interpretations of these are discussed. Importantly, the proposed algorithms preserve the states of subsystems in the whole-system model, so that analysis of root causes of instability can indicate specific states and parameters in the physical system. The theoretical advances are validated through numerical analysis and electromagnetic transient (EMT) simulation of example power systems.

The rest of this article is organized as follows: The fundamentals of state space and descriptor state space are briefly reviewed in \sectionref{Section:FundamentalDss}. In \sectionref{Section:AdvanceDss}, the calculation and interpretation of descriptor state space for power system applications are elaborated. \sectionref{Section:CaseStudy} gives the case studies with numerical results and EMT simulations. \sectionref{Section:Conclusions} concludes the article.

% ===========================================
% Fundamentals
% ===========================================

\section{Fundamentals of Descriptor State Space} \label{Section:FundamentalDss}

A standard state space model is \citeref{skogestad2005multivariable}
\begin{equation}    \label{Equ:Ss}
\begin{aligned}
& \dot{x} = Ax + Bu
\\
& y = Cx + Du
\end{aligned}
\end{equation}
where $x$, $u$, and $y$ are system vectors of state, input, and output, respectively; $A$, $B$, $C$, $D$ are system matrices. This state space system is equivalent to the transfer function system $G(s)$ as
\begin{equation}    \label{Equ:SsTf}
y = 
\underbrace{\big(C(sI - A)^{-1}B+D\big)}_{G(s)} 
u
\end{equation}
where $s$ is the Laplace operator and $I$ is an identity matrix. It is worth highlighting that, the system in \equref{Equ:Ss} and \equref{Equ:SsTf} is a proper system, in which the degree of numerator (order of zeros) does not exceed the degree of denominator (order of poles) \citeref{oppenheim1997signals,skogestad2005multivariable}. This means that the input $u$ and output $y$ can not be freely selected. For example, for a $s$-domain inductor, the inductor current is the system state which is also the system output, and the inductor voltage is the system input. The corresponding state space model is
\begin{equation} \label{Equ:InductorY}
\begin{aligned}
    \dot{i} 
    & = \underbrace{\begin{bmatrix} 0 \end{bmatrix}}_A i + \underbrace{\begin{bmatrix} 1/L \end{bmatrix}}_B v
    \\
    i 
    & = \underbrace{\begin{bmatrix} 1 \end{bmatrix}}_C i + \underbrace{\begin{bmatrix} 0 \end{bmatrix}}_D v
\end{aligned}
\end{equation}
This corresponds to an admittance model of $G(s) = 1/(sL)$ with the inductance $L$, i.e., a proper system.

For representing an improper system by state space method, the descriptor state space (also known as implicit state space) can be used, whose format is \citeref{skogestad2005multivariable}
\begin{equation}    \label{Equ:Dss}
\begin{aligned}
E\dot{x} & = Ax + Bu
\\
y & = Cx + Du 
\end{aligned}
\end{equation}
It is a generalized state space format with an additional matrix $E$. The corresponding transfer function representation of \equref{Equ:Dss} is
\begin{equation}    \label{Equ:DssTf}
y = 
\underbrace{\big(C(sE-A)^{-1}B+D\big)}_{G(s)} 
u
\end{equation}
Obviously, \equref{Equ:Dss} and \equref{Equ:DssTf} can also be used for modeling a proper system when $E$ is invertable or simply an identity matrix $I$, and we can easily convert \equref{Equ:Dss} to \equref{Equ:Ss} as
\begin{equation}    \label{Equ:Dss2SsInvE}
\begin{aligned}
\dot{x} & = [E^{-1}A] x + [E^{-1}B] u
\\
y & = Cx + Du 
\end{aligned}
\end{equation}
By contrast, when \equref{Equ:Dss} is an improper system, $E$ is a singular matrix, i.e., not invertable and not full-rank \citeref{skogestad2005multivariable}. In this case, \equref{Equ:Dss} cannot be converted to \equref{Equ:Ss}. Again, a $s$-domain inductor is taken as an example here, whose impedance model $G(s) = sL$ is improper. Its corresponding descriptor state space representation is
\begin{equation} \label{Equ:InductorZ}
\begin{aligned}
    \underbrace{\begin{bmatrix} L & 0 \\ 0 & 0 \end{bmatrix}}_E \begin{bmatrix} \dot{i} \\ \dot{v} \end{bmatrix}
    & = 
    \underbrace{\begin{bmatrix} 0 & 1 \\ -1 & 0 \end{bmatrix}}_A \begin{bmatrix} i \\ v \end{bmatrix}
    + 
    \underbrace{\begin{bmatrix} 0 \\ 1 \end{bmatrix}}_B i
    \\
    v 
    & =  
    \underbrace{\begin{bmatrix} 0 & 1 \end{bmatrix}}_C \begin{bmatrix} i \\ v \end{bmatrix}
    +
    \underbrace{\begin{bmatrix} 0 \end{bmatrix}}_D i
\end{aligned}
\end{equation}
where $E$ is not invertable.

\begin{figure*}[t!]
\centering
\includegraphics[scale=0.95]{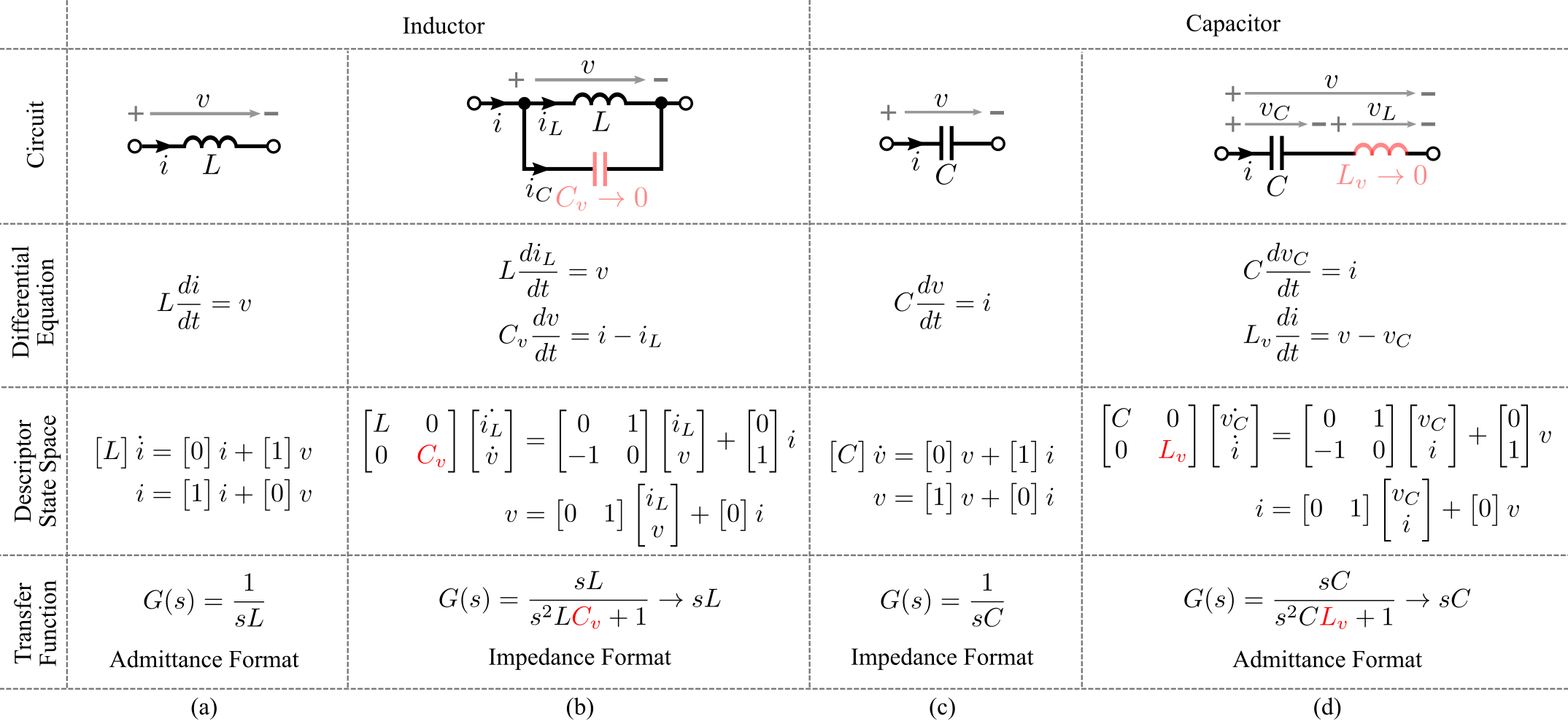}
\caption{Descriptor state space model of fundamental electrical circuit elements. (a) Admittance model of an inductor. (b) Impedance model of an inductor. (c) Impedance model of a capacitor. (d) Admittance model of a capacitor.}
\label{Fig:DssInductorCapacitor}
\end{figure*}

% ===========================================
% Advance
% ===========================================

\section{Calculation and Interpretation of Descriptor State Space} \label{Section:AdvanceDss}

When modularly modeling a large power system, combing subsystems together normally needs the connection of subsystem models. This further leads to the necessity and advantages of using descriptor state space over standard state space or transfer function, as discussed in this section. Calculation algorithms of using descriptor state space are also derived next.

\subsection{Inverse of Descriptor State Space}

For a standard state space model in \equref{Equ:Ss}, when $D$ is invertable, its inverse can be easily obtained as \citeref{skogestad2005multivariable}
\begin{equation}
\begin{aligned}
& \dot{x} = [A - B D^{-1} C] x + [D^{-1}] y
\\
& u = [- D^{-1} C] x + [D^{-1}] y
\end{aligned}
\end{equation}
However, an invertable $D$ matrix means the system has the same dimension of input $u$ and output $y$ and the equivalent system transfer function has the same the degrees of numerator and denominator, i.e., proper but not strictly proper \citeref{skogestad2005multivariable}. This is a very strict condition and is always not satisfied in power systems. Counter examples can be easily found when looking at models of inductors and capacitors. Therefore, a generalized inverse of \equref{Equ:Dss} is proposed next. We firstly re-write the descriptor state space model in \equref{Equ:Dss} as
\begin{equation}
\begin{aligned}
E\dot{x} & = Ax+Bu+0y
\\
0\dot{u} & = -Cx-Du+y
\end{aligned}
\end{equation}
Then, if regarding this equation as the new state equation ($x$ and $u$ are states and $y$ is input), we can get the system model as
\begin{equation} \label{Equ:DssInverse}
    \begin{aligned}
    \begin{bmatrix}  E & 0 \\ 0 & 0 \end{bmatrix} \begin{bmatrix} \dot{x} \\ \dot{u} \end{bmatrix}
    & =
    \underbrace{\begin{bmatrix}  A & B \\ -C & -D \end{bmatrix}}_{\text{New}~A} 
    \begin{bmatrix} x \\ u \end{bmatrix}
    +
    \underbrace{\begin{bmatrix} 0 \\ I \end{bmatrix}}_{\text{New}~B} y
    \\
    u 
    & =
    \underbrace{\begin{bmatrix} 0 & I \end{bmatrix}}_{\text{New}~C} 
    \begin{bmatrix} x \\ u \end{bmatrix}
    +
    \underbrace{\begin{bmatrix} 0 \end{bmatrix}}_{\text{New}~D} y
    \end{aligned}
\end{equation}
which is an inverse system of \equref{Equ:Dss} with new input $y$ and new output $u$. Remarkably, $u$ is also added into the state vector as a new virtual state, which is the key of achieving the model inverse. Next, inductors and capacitors will be used to physically interpret this inverse algorithm.

%Figure (a) shows a full inverse of input $u$ and output $y$, and (b) gives a general inverse of only swapping partial input $u_1$ and partial output $y_1$ but keeping $u_2$ and $y_2$ unchanged. The key step of inverse is to introduce a new virtual state $\xi$ which equals to the old system input. Next, an inductor and a capacitor will be used to physically interpret the inverse of a state space model.

\figref{Fig:DssInductorCapacitor}(a) shows the state space representation of an admittance model of an inductor, i.e., $1/(sL)$ a strictly proper system. According to \equref{Equ:DssInverse}, the key of obtaining the inverse model needs to introduce a new virtual state which is exactly the system old input, i.e., inductor voltage. Hence, a virtual capacitor $C_v$ is added in parallel with the inductor $L$. This leads to the new model in \figref{Fig:DssInductorCapacitor}(b). Remarkably, when $C_v$ is ideally zero, we can get $i_C = 0$, $i_L=i$, and a singular matrix $E=[L,0;0,0]$. In this case, the model in \figref{Fig:DssInductorCapacitor}(b) is essentially the descriptor state space representation of $sL$, i.e., an improper system. The model in \figref{Fig:DssInductorCapacitor}(b) exactly coincide with \equref{Equ:InductorZ} in last section and the inverse algorithm in \equref{Equ:DssInverse}. Similarly, for a capacitor, \figref{Fig:DssInductorCapacitor}(c) shows the impedance model $1/(sC)$, and \figref{Fig:DssInductorCapacitor}(d) shows the admittance model $sC$ by adding a virtual inductor of $L_v=0$. It is also worth highlighting that, when $C_v$ and $L_v$ are ideally zero, the systems in \figref{Fig:DssInductorCapacitor}(b) and (d) can not be represented by conventional state space anymore because of the non-invertable $E$ matrix.

In conclusion, the physical interpretation of the inverse of an electrical impedance or admittance in state space is: a new virtual state of voltage or current is added by adding a virtual capacitor of $C_v=0$ in parallel or a virtual inductor of $L_v=0$ in series. 

\subsection{Connection of Descriptor State Space} \label{Section:Connection}

\begin{figure*}[t!]
\centering
\includegraphics[scale=0.7]{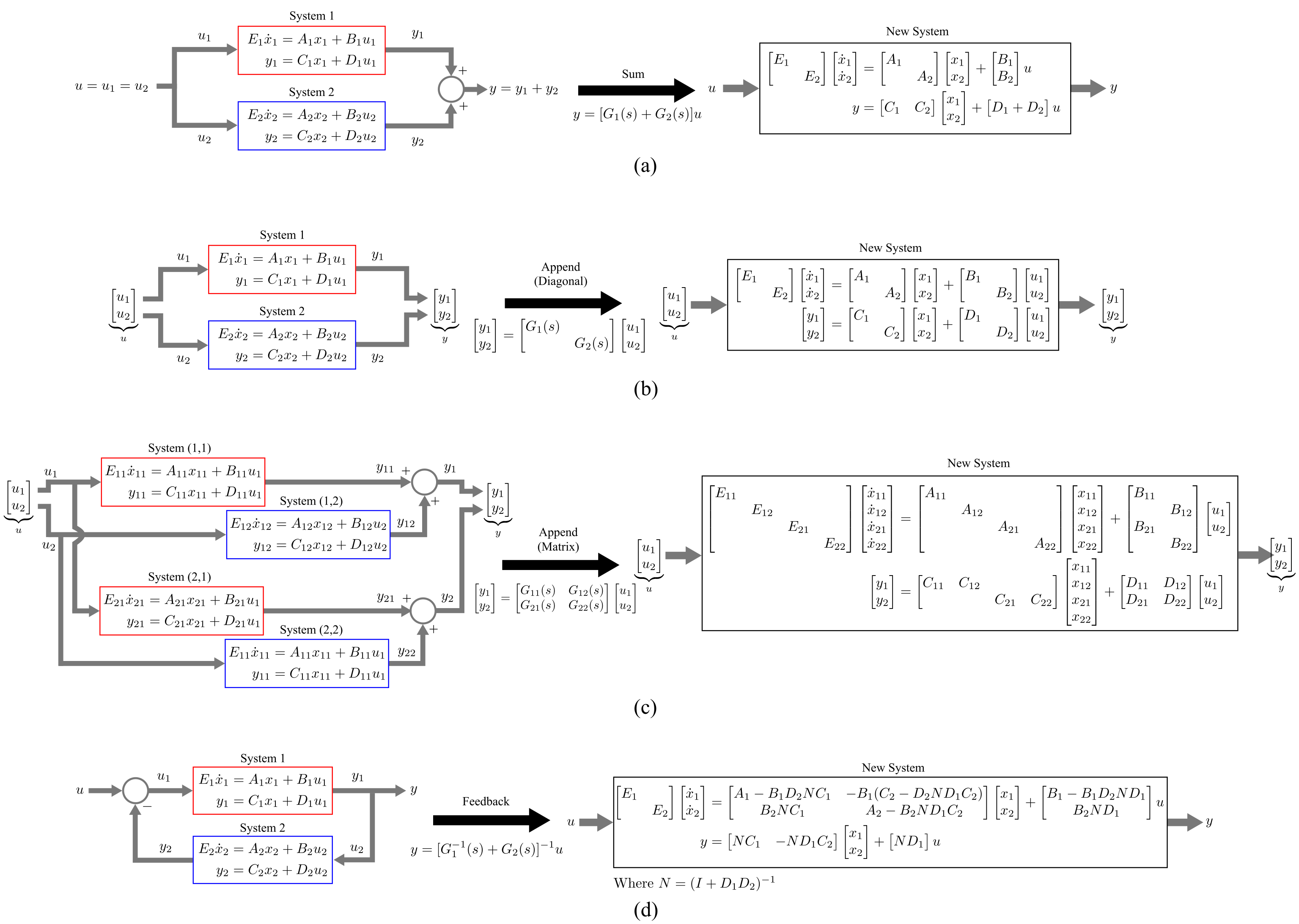}
\caption{Algorithms of combining and connecting descriptor state space models. (a) Sum. (b) Append. (c) Matrix append. (d) Feedback.}
\label{Fig:CalculationRule}
\end{figure*}

\begin{figure*}[t!]
\centering
\includegraphics[scale=0.85]{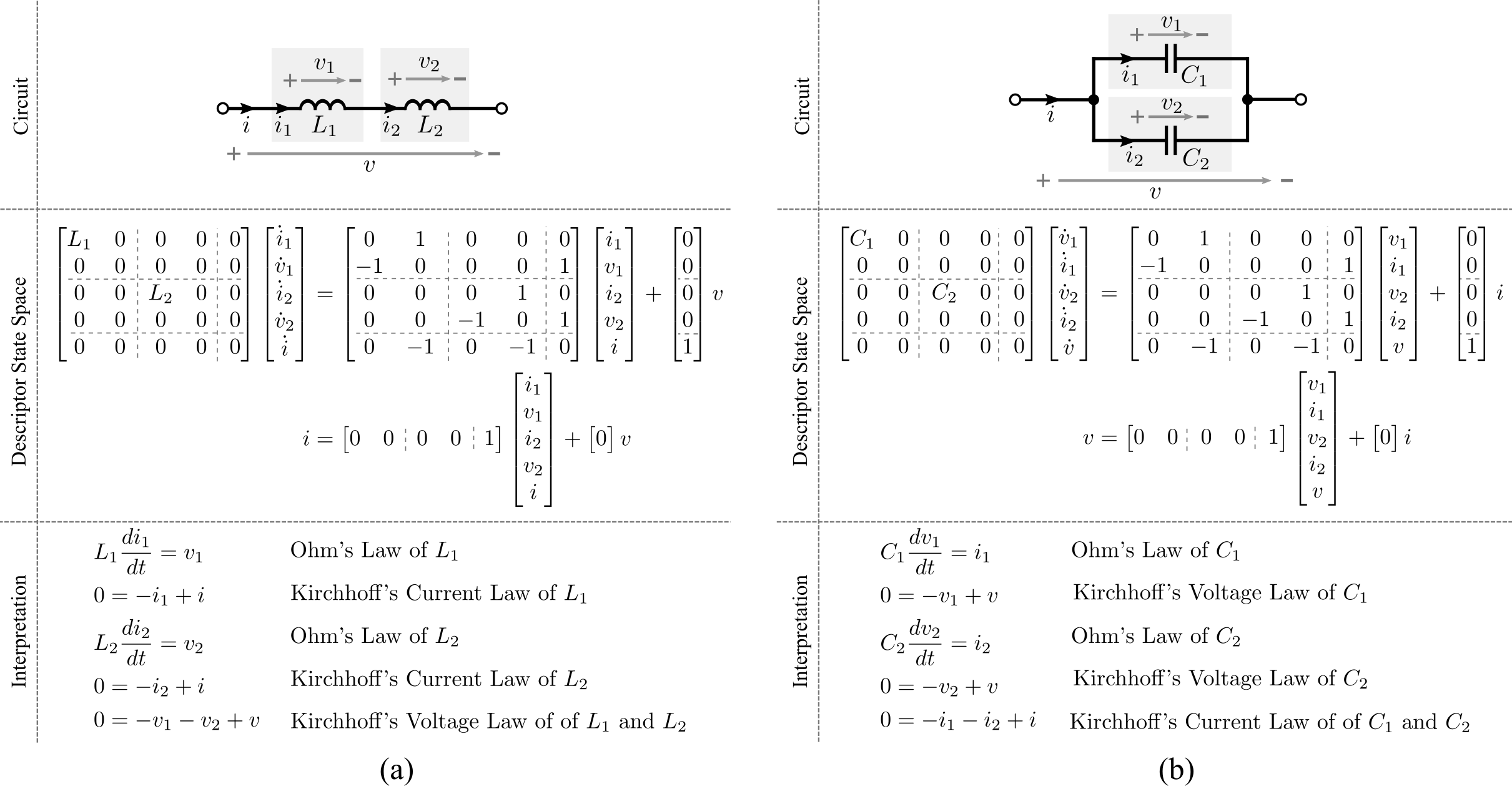}
\caption{Modeling the connection of improper systems. (a) Series connection of inductors. (b) Parallel connection of capacitors.}
\label{Fig:ImproperConnection}
\end{figure*}

When modeling a large power system in a modular way, the models of subsystems are derived first, and then the whole system model is obtained by connecting subsystem models together. However, as discussed in \sectionref{Section:Introduction}, subsystems in practical power systems are normally ``terminal-inductive''. This leads to the model port mismatch between subsystems and sufficiently-large virtual resistors \citeref{pogaku2007modeling,wang2017small} or sufficiently-small virtual capacitors \citeref{markovic2019understanding,PSCAD} are added to solve this problem, which change the system physical structure, introduce errors, and limit the modeling flexibility.

The underlying reason of the above mentioned port mismatch problem is essentially because standard state space cannot represent the connection of improper systems, e.g., the series connection of inductors or parallel connection of capacitors. By contrast, the descriptor state space can naturally solve this problem. In addition to the inverse algorithm in \equref{Equ:DssInverse}, other commonly used algorithms (sum, append, feedback, etc) of connecting descriptor state space are summarized in \figref{Fig:CalculationRule}. The detailed derivation of these algorithms is similar to that of standard state space in \citeref{duke1986combing} and is omitted here. Next, an example of modeling the series connection of two inductors is investigated. The descriptor state space model is illustrated in \figref{Fig:ImproperConnection}(a), which can be obtained by modeling
\newline \textit{Step 1}: $1/(sL_1)$ and $1/(sL_2)$ via standard state space;
\newline \textit{Step 2}: $sL_1$ and $sL_2$ via inverse in \equref{Equ:DssInverse};
\newline \textit{Step 3}: $(sL_1 + sL_2)$ via sum in \figref{Fig:CalculationRule}(a);
\newline \textit{Step 4}: $1/(sL_1 + sL_2)$ via inverse in \equref{Equ:DssInverse}.
\newline which gives the final model. The model correctness can be easily validated by deriving the corresponding transfer function according to \equref{Equ:DssTf}. As summarized in \figref{Fig:ImproperConnection}(a), each state equation in the model has a clear physical interpretation, such as Ohm's law, Kirchhoff's voltage law, Kirchhoff's current law, etc. For example, three Kirchhoff equations clearly indicate the series connection relation of two inductors. Additionally, these three Kirchhoff equations are also state equations of $v_1$, $v_2$, and $i$, which are associated to three virtual circuit elements: $v_1$ is the state of a virtual capacitor in parallel with $L_1$, given by step 2; $v_2$ is the state of a virtual capacitor in parallel with $L_2$, given by step 2; $i$ is the state of a virtual inductor in series with the whole model, given by step 4. Similarly to modeling inductors, \figref{Fig:ImproperConnection}(b) shows the dual example of modeling the parallel connection of two capacitors. 

In general, according to the calculation algorithms, we can get the model of a power system without introducing modeling errors and restricting the choice of system input or output variables. For instance, 
\newline \textit{Step1}: Similarly to the derivation of nodal admittance matrix \citeref{kundur1994power,gu2021impedance,li2022mapping}, the whole-network model can be derived by combing the model of each branch, via algorithms of sum and matrix append in \figref{Fig:CalculationRule}(a) and (c).
\newline \textit{Step 2}: Models of all apparatuses (IBRs, SGs, etc) in a power grid can also be combined via the append in \figref{Fig:CalculationRule}(b), which gives the whole-apparatus model.
\newline \textit{Step 3}: The whole-network model (Step 1) and whole-apparatus model (Step 2) can be connected via the feedback rule in \figref{Fig:CalculationRule}(d), which gives the so-called whole-system impedance model \citeref{gu2021impedance,zhu2022participation,zhuEAimpedance} or port-mapping model \citeref{li2022mapping}.
\newline The whole-system model can then be used for analyzing system dynamics, tuning parameters, and enhancing stability.

\subsection{Transforming Descriptor State Space to State Space} \label{Section:Conversion}

As mentioned in \sectionref{Section:FundamentalDss}, a physically-realizable system must be proper because of causalty \citeref{oppenheim1997signals}. Even though a proper system can be represented by descriptor state space, it can also be transformed back to an equivalent state space model. The algorithm for that is now derived.

For a descriptor state space system in \equref{Equ:Dss}, if $E$ is invertable, then the conversion can be easily done by left-multiplying $E^{-1}$ to the state equation, as shown in \equref{Equ:Dss2SsInvE}. Examples of this case can be found in \figref{Fig:DssInductorCapacitor}(a) and (c). If $E$ is not invertable, we firstly re-write \equref{Equ:Dss} as
\begin{equation}    \label{Equ:DssE1}
    \begin{aligned}
        \begin{bmatrix} E_1 & 0 \\ 0 & 0 \end{bmatrix}
        \begin{bmatrix} \dot{x}_1 \\ \dot{x}_2 \end{bmatrix}
        & =
        \begin{bmatrix} A_{11} & A_{12} \\ A_{21} & A_{22} \end{bmatrix}
        \begin{bmatrix} x_1 \\ x_2 \end{bmatrix}
        +
        \begin{bmatrix} B_1 \\ B_2 \end{bmatrix}
        u
        \\
        y
        & =
        \begin{bmatrix} C_1 & C_2 \end{bmatrix}
        \begin{bmatrix} x_1 \\ x_2 \end{bmatrix}
        +
        \begin{bmatrix} D \end{bmatrix} u
    \end{aligned}
\end{equation}
where $E_1$ is invertable, $x_1$ consists of practical states, and $x_2$ consists of virtual states. Examples of this case can be found in \figref{Fig:ImproperConnection}(a) and (b), where $E_1=[L_1,0;0,L_2]$ and $x_1=[i_1;i_2]$ for (a) and $E_1=[C_1,0;0,C_2]$ and $x_1=[v_1;v_2]$ for (b). The next key step of achieving the transform is to find $x_2 = f(x_1,u_1)$ so that we can eliminate the virtual state $x_2$ in \equref{Equ:DssE1}. There are two sub-cases, as elaborated next.

When $A_{22}$ is invertable, $x_2$ can be directly solved as
\begin{equation}
    x_2 = -A_{22}^{-1}A_{21} x_1 - A_{22}^{-1}B_2 u
\end{equation}
Then using this equation to replace $x_2$ in \equref{Equ:DssE1} yields the state space representation as
\small
\begin{equation}    \label{Equ:Dss2SsInvA22}
\begin{aligned}
    \dot{x_1} 
    = & 
    \underbrace{\begin{bmatrix} E_1^{-1}(A_{11}-A_{12}A_{22}^{-1}A_{21}) \end{bmatrix}}_{\text{New}~A} x_1
    +
    \underbrace{\begin{bmatrix} E_1^{-1}(B_1 - A_{12} A_{22}^{-1}B_2) \end{bmatrix}}_{\text{New}~B} u
    \\
    y 
    = &
    \underbrace{\begin{bmatrix} C_1-C_2A_{22}^{-1}A_{21} \end{bmatrix}}_{\text{New}~C} x_1
    +
    \underbrace{\begin{bmatrix} D-C_2A_{22}^{-1}B_2 \end{bmatrix}}_{\text{New}~D} u
\end{aligned}
\end{equation}

When $A_{22}$ is not invertable, we first find the null matrix $N$ of $A_{22}$ so that \citeref{axler1997linear}
\begin{equation} \label{Equ:NullA22}
    \begin{aligned}
        NA_{22} = 0,~\text{Rank}(N)+\text{Rank}(A_{22}) = \text{Order}(A_{22})
    \end{aligned}
\end{equation}
Then, left-multiplying $N$ to the second state equation in \equref{Equ:DssE1} yields
\begin{equation}
\begin{aligned}
0 = N A_{21} x_1 + \underbrace{NA_{22}}_0 x_2 + N B_2 u
\end{aligned}
\end{equation}
The time derivative of this equation is
\begin{equation}
0 = N A_{21} \dot{x}_1 + N B_2 \dot{u}
\end{equation}
Replacing $\dot{x}_1$ in this equation by using the first state equation in \equref{Equ:DssE1}, we get
\begin{equation}
\begin{aligned}
0 = \underbrace{N A_{21}E_1^{-1}}_{\hat{N}} (A_{11}x_1 + A_{12}x_2 + B_1 u) + N B_2 \dot{u}
\end{aligned}
\end{equation}
Combining this equation with the second state equation in \equref{Equ:DssE1}, we get the augmented equation as
\begin{equation} 
    0
    = 
    \underbrace{\begin{bmatrix} A_{21} \\ \hat{N} A_{11} \end{bmatrix}}_{\hat{A}_{21}} x_1 
    +
    \underbrace{\begin{bmatrix} A_{22} \\ \hat{N} A_{22} \end{bmatrix}}_{\hat{A}_{22}} x_2
    +
    \underbrace{\begin{bmatrix} B_2 \\ \hat{N} B_1 \end{bmatrix}}_{\hat{B}_2} u
    +
    \underbrace{\begin{bmatrix} 0 \\ \hat{N} B_2 \end{bmatrix}}_{\hat{B}_{d}} \dot{u}
\end{equation}
where $\hat{A}_{21}$, $\hat{A}_{22}$, $\hat{B}_2$, and $\hat{B}_d$ are the augmented matrices. It is remarkable that $\hat{A}_{22}$ is generalized-invertable (specifically, left-invertable) because it consists of $A_{22}$ and its null matrix $N$ \citeref{axler1997linear}. Now, $x_2$ can be solved as
\begin{equation}
    x_2 
    = 
    - \hat{A}_{22}^{-1} \hat{A}_{21} x_1
    - \hat{A}_{22}^{-1} \hat{B}_2 u
    - \hat{A}_{22}^{-1} \hat{B}_d \dot{u}
\end{equation}
Using this equation to replace $x_2$ in \equref{Equ:DssE1}, we can get the state space representation as
\small
\begin{equation} \label{Equ:Dss2SsInvA22Null}
\begin{aligned}
    \dot{x}_1 
    = & 
    \underbrace{\begin{bmatrix} E_1^{-1}(A_{11} - A_{12} \hat{A}_{22}^{-1} \hat{A}_{21}) \end{bmatrix}}_{\text{New}~A} x_1 
    +
    \underbrace{\begin{bmatrix} E_1^{-1}(B_1-A_{12} \hat{A}_{22}^{-1} \hat{B}_2) \end{bmatrix}}_{\text{New}~B} u
    \\
    & +
    \underbrace{\begin{bmatrix} -E_1^{-1}A_{12} \hat{A}_{22}^{-1} \hat{B}_d \end{bmatrix}}_{B_d} \dot{u}
    \\
    y
    = &
    \underbrace{\begin{bmatrix} C_1 - C_2 \hat{A}_{22}^{-1}\hat{A}_{21} \end{bmatrix}}_{\text{New}~C} x_1
    +
    \underbrace{\begin{bmatrix} D - C_2\hat{A}_{22}^{-1}\hat{B}_2) \end{bmatrix}}_{\text{New}~D} u
    \\
    & +
    \underbrace{\begin{bmatrix} -C_2 \hat{A}_{22}^{-1} \hat{B}_{d} \end{bmatrix}}_{D_d} \dot{u}
\end{aligned}
\end{equation}
We can easily see that \equref{Equ:Dss2SsInvA22} is simply a special case of \equref{Equ:Dss2SsInvA22Null} when $A_{22}$ is full-rank and its null matrix $N$ is empty, and therefore $\hat{A}_{21}=A_{21}$, $\hat{A}_{22}=A_{22}$, $\hat{B}_{2}=B_{2}$. It is also worth mentioning that, the additional matrices $B_d$ and $D_d$ are non-zeros only for improper system because of its higher order of zeros than poles (the time derivative of input $u$). As for causal and proper systems in practice, $B_d$ and $D_d$ are zero. 

A simple example is also given here. The admittance model of two series-connected inductors in \figref{Fig:ImproperConnection}(a) is a proper system that represented by descriptor state space. By using \equref{Equ:Dss2SsInvA22Null}, this model can be converted back to state space representation as
\begin{equation}
    \begin{aligned}
    \begin{bmatrix} \dot{i}_1 \\ \dot{i}_2 \end{bmatrix} 
    & = 
    \underbrace{\begin{bmatrix} 0 & 0 \\ 0 & 0 \end{bmatrix}}_{\text{New} A} \begin{bmatrix} i_1 \\ i_2 \end{bmatrix}
    +
    \underbrace{\begin{bmatrix} 1/(L_1+L_2) \\ 1/(L_1+L_2) \end{bmatrix}}_{\text{New} B} v
    +
    \underbrace{\begin{bmatrix} 0 \\ 0 \end{bmatrix}}_{B_d} \dot{v}
    \\
    i
    & =
    \underbrace{\begin{bmatrix} 1/2 & 1/2 \end{bmatrix}}_{\text{New} C} \begin{bmatrix} i_1 \\ i_2 \end{bmatrix}
    +
    \underbrace{\begin{bmatrix} 0 \end{bmatrix}}_{\text{New} D} v
    +
    \underbrace{\begin{bmatrix} 0 \end{bmatrix}}_{D_d} \dot{v}
    \end{aligned}
\end{equation}
which can be further simplified to
\begin{equation}
    \begin{aligned}
    \dot{i}
    & = 
    \underbrace{\begin{bmatrix} 0 \end{bmatrix}}_{\text{New}~A} i
    +
    \underbrace{\begin{bmatrix} 1/(L_1+L_2) \end{bmatrix}}_{\text{New}~B} v
    \\
    i
    & =
    \underbrace{\begin{bmatrix} 1 \end{bmatrix}}_{\text{New}~C} i
    +
    \underbrace{\begin{bmatrix} 0 \end{bmatrix}}_{\text{New}~D} v
    \end{aligned}
\end{equation}
i.e., a standard state space model of $1/[s(L_1+L_2)]$.

\subsection{Tracked States in Descriptor State Space}

It is worth highlighting that, the system states are always tracked in the proposed algorithm. For example, in model inverse algorithm in \equref{Equ:DssInverse}, the new system state consists of the old system state with an additional virtual state $u$; in model transform algorithm in \equref{Equ:Dss2SsInvA22Null}, the new system state is the subset of old system state; in model connection algorithm in \figref{Fig:CalculationRule}, the new system state is the union set of subsystem states. In other words, in the finally obtained whole system model, a state can always be mapped to a physical state in certain subsystem. This enables investigating the participation of a physical state on system oscillation modes, and can be used for analyzing system dynamics and tuning system parameters, as will be illustrated by case studies in next section. The participation analysis can be done either in the converted state space format in \equref{Equ:Dss2SsInvA22Null}, or in the descriptor state space format directly as derived in \appendixref{Appendix:Participation}.

% A toolbox is developed based on the descriptor state space for the modeling and dynamic analysis of multi-bus large-scale power systems, which is open-source online \citeref{FuturePowerNetworks}.

% ===========================================
% Case Study
% ===========================================

\section{Case Study} \label{Section:CaseStudy}

\begin{figure}[t!]
\centering
\includegraphics[scale=0.8]{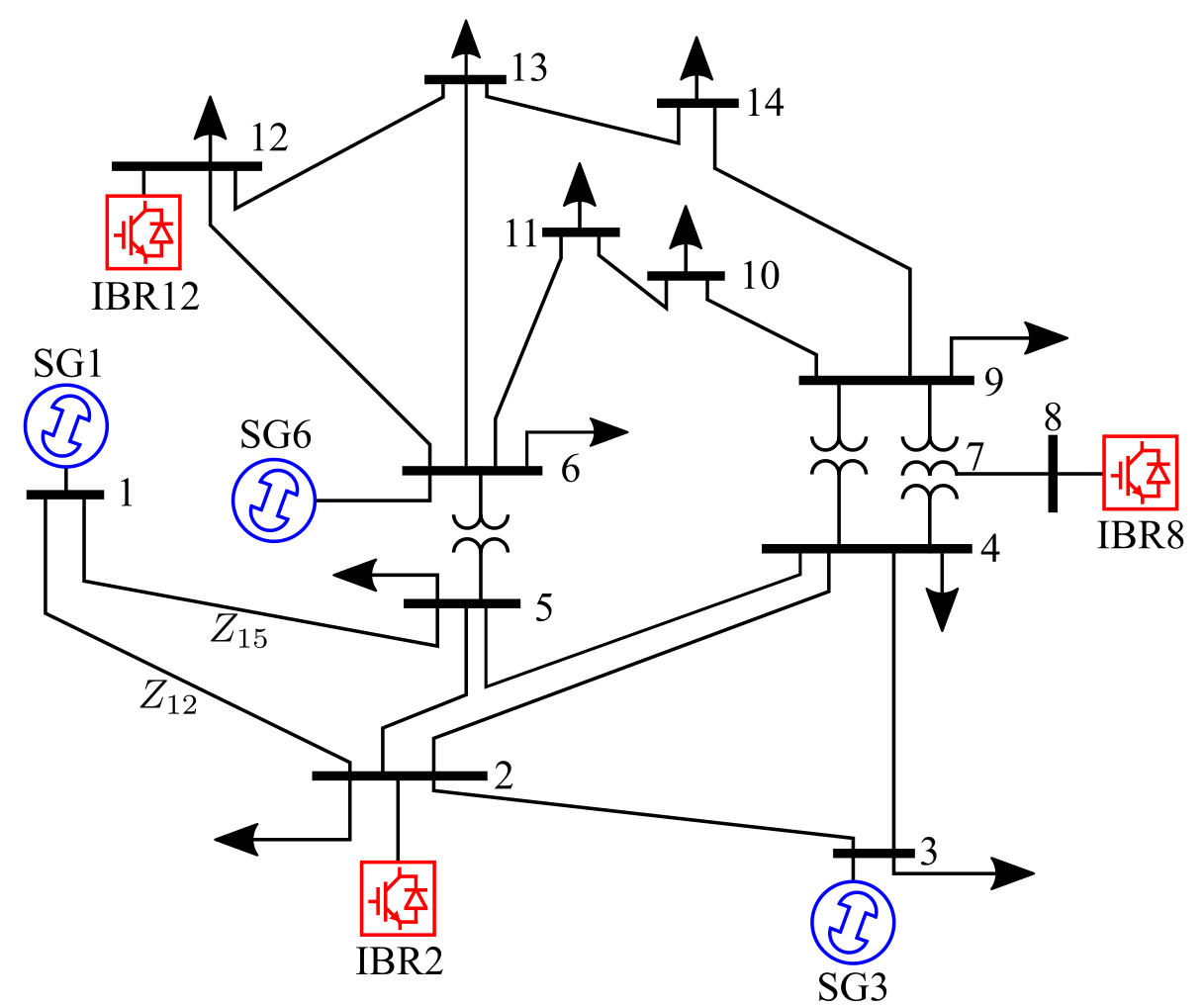}
\caption{A modified IEEE 14-bus power system.}
\label{Fig:14Bus}
\end{figure}

\begin{figure}[t!]
\centering
\includegraphics[scale=0.8]{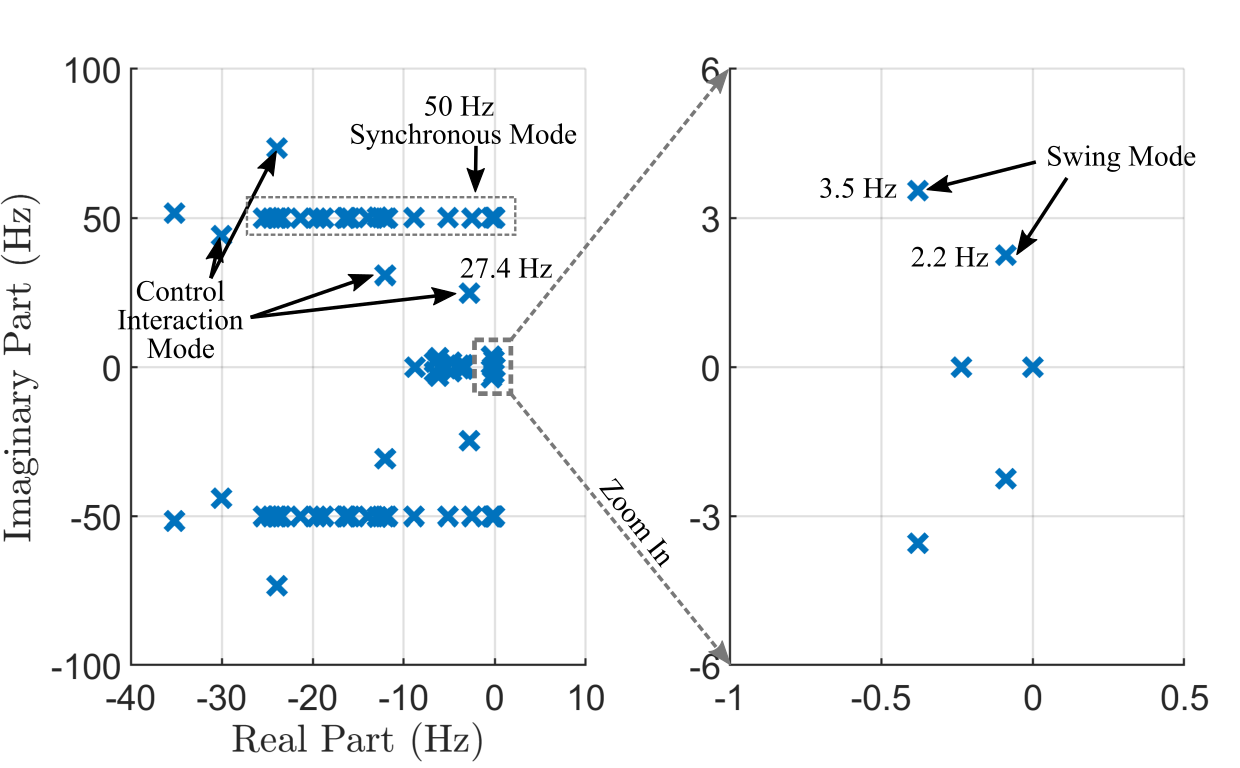}
\caption{Whole-system mode map.}
\label{Fig:ModeMap}
\end{figure}

\begin{figure}[t!]
\centering
\includegraphics[scale=0.65]{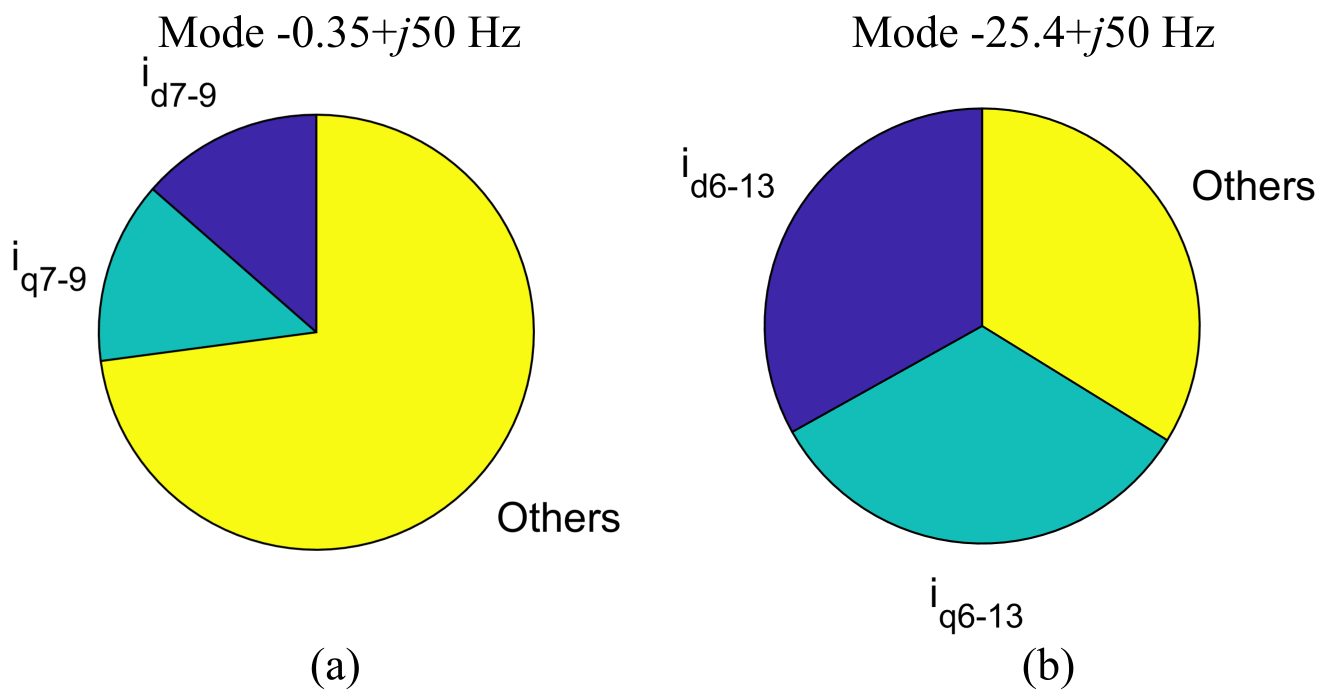}
\caption{Synchronous mode analysis of lines. (a) Participation on -0.35+$j$50 Hz mode. (b) Participation on -25.4+$j$50 Hz mode.}
\label{Fig:50HzMode}
\end{figure}

\begin{figure}[t!]
\centering
\includegraphics[scale=0.65]{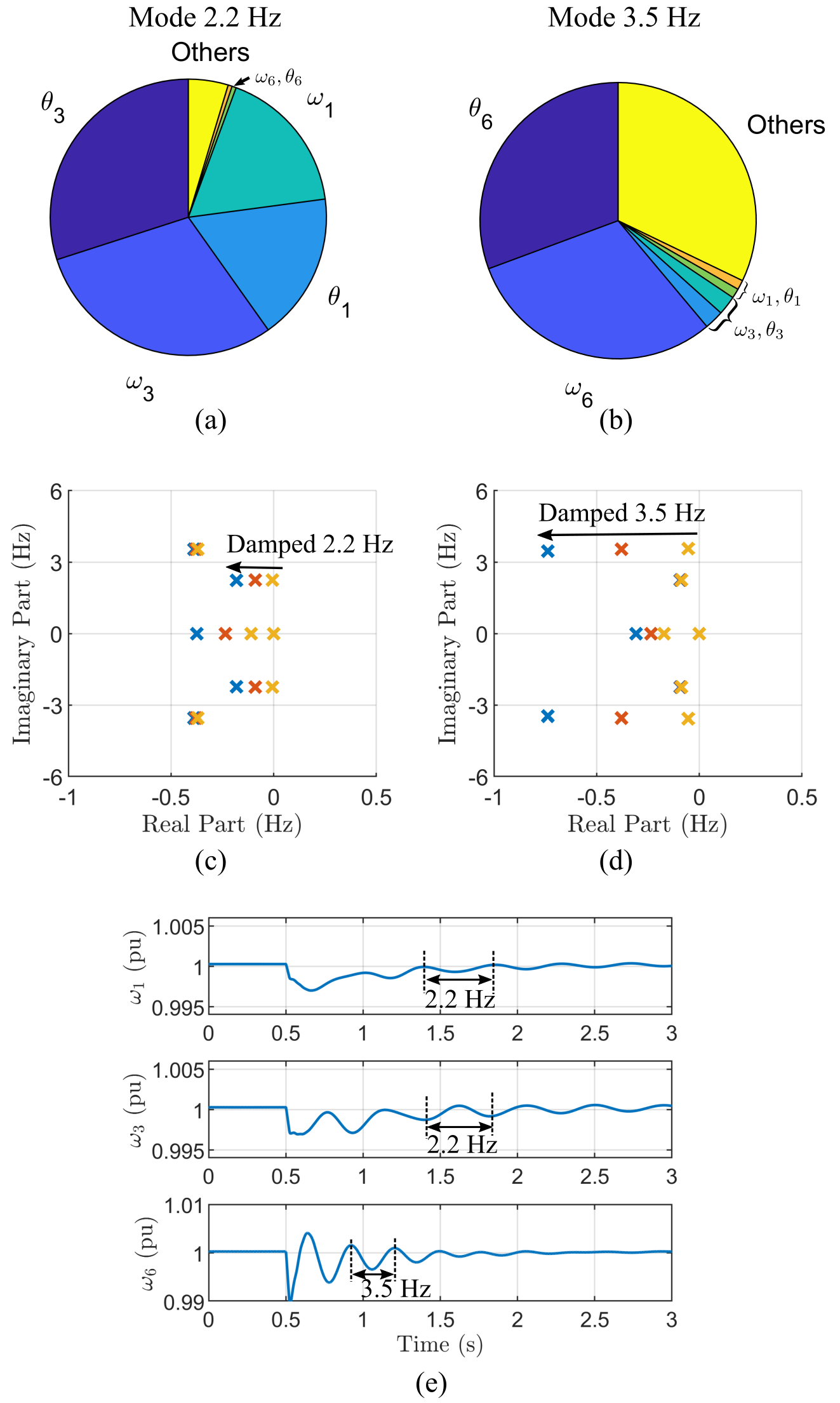}
\caption{Swing mode analysis of SGs. (a) Participation on 2.2 Hz mode. (b) Participation on 3.5 Hz mode. (c) Mode locus with increasing rotor damping of SG1 and SG3 from 1 pu to 20 pu. (d) Mode locus with increasing rotor damping of SG6 from 1 pu to 20 pu. (e) EMT simulation results with 1 pu load impulse at all SG buses at 0.5 s.}
\label{Fig:SwingMode}
\end{figure}

By using descriptor state space method proposed in last section, a modified IEEE 14-bus system in \figref{Fig:14Bus} is modeled numerically. The system consists of 3 SGs and 3 IBRs (Type-IV wind generation). The parameters, scripts, and models are open-source online \citeref{FuturePowerNetworks} and are integrated into a toolbox \citeref{SimplusGT}. The system dynamics are analyzed by eigenvalue and state participation (descriptor state space model) and validated by time-domain simulations (EMT model in Matlab/Simulink) next. The eigenvalues (also known as modes) of the whole-system are shown in \figref{Fig:ModeMap}. Certain modes are highlighted and analyzed in detail next.

\subsection{Synchronous Mode of Inductance}

The 50 Hz synchronous mode in the mode map of \figref{Fig:ModeMap} is investigated first. \figref{Fig:50HzMode} shows the participation analysis on two modes with different real parts: (a) $-0.35+j50$ Hz; and (b) $-25.4+j50$ Hz. The participation pie chart shows that mode (a) has a significant participation from branch 7-9 which is a transformer winding with very large X/R ratio of 20, that is, a component with low damping. This is reflected in the mode being close to imaginary axis. In contrast, mode (b) is dominated by branch 6-13 which is a line with X/R ratio of $1.97 = 0.13027/0.06615$, i.e., properly damped. This is reflected in this mode being located far from the imaginary axis.

\subsection{Swing mode of SG}

\begin{figure*}[th!]
\centering
\includegraphics[scale=0.75]{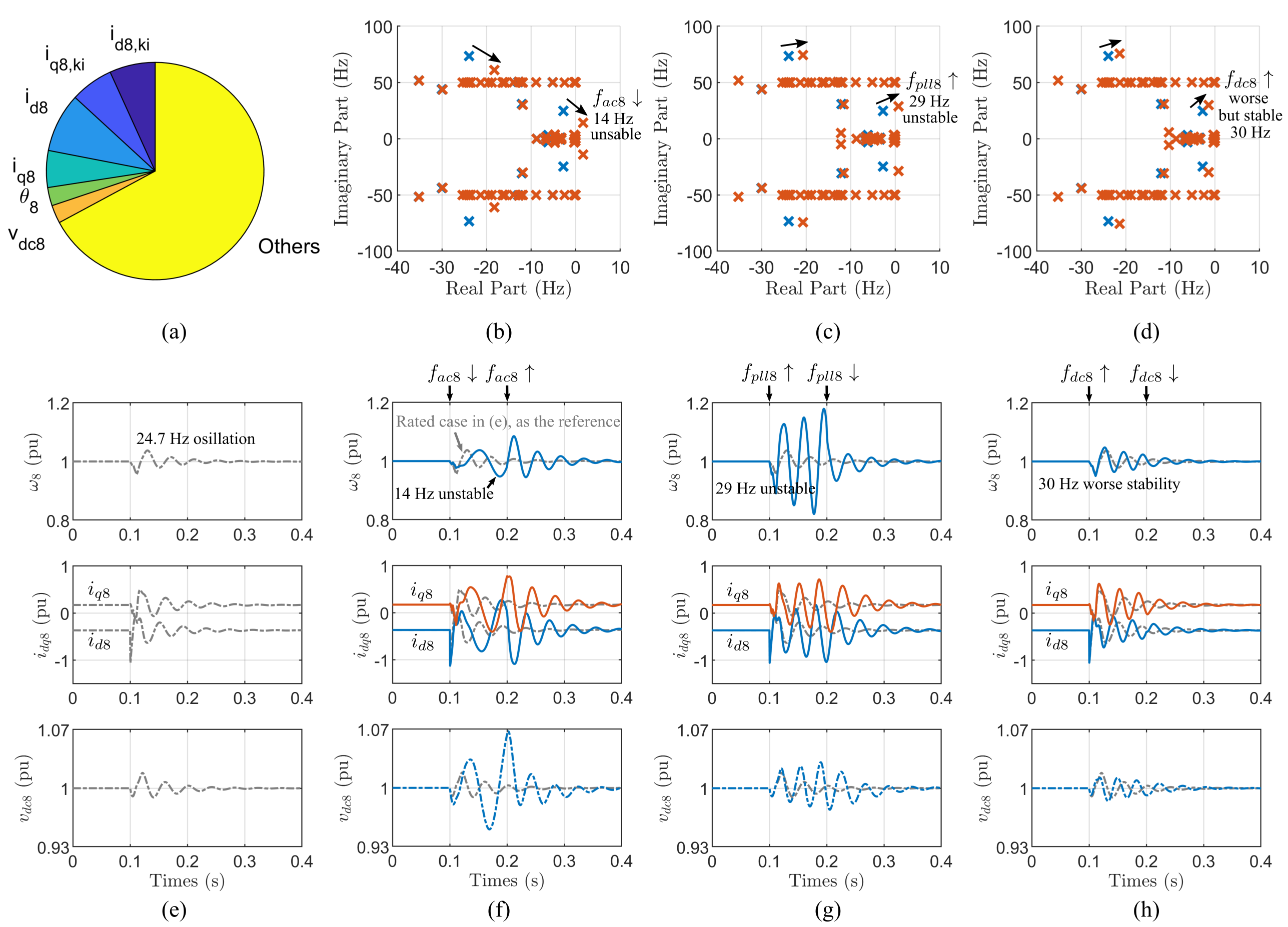}
\caption{Control interaction mode analysis of IBR8. (a) Participation on 24.7 Hz mode. (b) Mode locus with reducing ideal current bandwidth $f_{ac8}$ from 250 Hz to 150 Hz. (c) Mode locus with increasing PLL bandwidth $f_{pll8}$ from 10 Hz to 30 Hz. (d) Mode locus with increasing dc-link bandwidth $f_{dc8}$ from 10 Hz to 30 Hz. (e) EMT simulation with rated parameters of IBR8. (f) EMT simulation with reducing $f_{ac8}$ from 250 Hz to 150 Hz at 0.1 s and increasing it back at 0.2 s. (g) EMT simulation with increasing $f_{pll8}$ from 10 Hz to 30 Hz at 0.1 s and reducing it back at 0.2 s. (h) EMT simulation with increasing $f_{dc8}$ from 10 Hz to 30 Hz at 0.1 s and reducing it back at 0.2 s.}
\label{Fig:InteractionMode}
\end{figure*}

The 2.2 Hz and 3.5 Hz swing modes in \figref{Fig:ModeMap} are investigated and the results are summarized in \figref{Fig:SwingMode}. As shown in \figref{Fig:SwingMode}(a), the 2.2 Hz mode is dominated by the swing of SG1 ($\omega_1$, $\theta_1$) and SG3 ($\omega_3$, $\theta_3$); As displayed in \figref{Fig:SwingMode}(b), the 3.5 Hz mode is dominated by the swing of SG6 ($\omega_6$, $\theta_6$). This coincides with the mode loci in \figref{Fig:SwingMode}(c) and (d): The 2.2 Hz mode is damped only when increasing the rotor damping of SG1 and SG3 in (c); and the 3.5 Hz mode is damped only when increasing the rotor damping of SG6 in (d). \figref{Fig:SwingMode}(e) shows the EMT simulation results. After the 1 pu load impulse at 0.5 s, SG1 and SG3 oscillates dominantly at 2.2 Hz, and SG6 oscillates at 3.5 Hz. This coincides with the numerical analysis of the descriptor state space model.

\subsection{Control Interaction Mode of IBR}

The 24.7 Hz mode in the mode map \figref{Fig:ModeMap} is investigated next. According to \figref{Fig:InteractionMode}(a), this mode is mainly participated by the ac current control of IBR8 ($i_{d8,ki}$, $i_{q8,ki}$, $i_{d8}$, $i_{q8}$), but also jointly participated by its phase-locked loop (PLL) ($\theta_8$) and dc-link control ($v_{dc8}$). In other words, it is an interaction mode of different control loops of IBR8. \figref{Fig:InteractionMode}(b), (c), and (d) show the mode loci with changing corresponding control parameters, which indicates that slower current control bandwidth in (b), or faster PLL in (c), or faster dc-link control in (d) can lead to more chances of interaction and instability. \figref{Fig:InteractionMode}(e) shows the EMT simulation of 1 pu load impulse response of IBR8 with rated parameters, as the simulation reference. The 24.7 Hz oscillation can be clearly observed. \figref{Fig:InteractionMode}(f), (g), and (h) show the EMT simulations with badly tuning ac current bandwidth $f_{ac8}$, PLL bandwidth $f_{pll8}$, and dc-link bandwidth $f_{dc8}$ at 0.1 s to de-stabilize the system, and tuning them back at 0.2 s to re-stabilize the system. Remarkably, (f) shows 14 Hz unstable oscillations, (g) shows 29 Hz unstable oscillations, whereas (h) shows worse but still stable oscillations compared to the rated case in (e). All EMT simulation results properly coincide with the mode maps predicted by the descriptor state space model in (a) to (d).
% \footnote{The current bandwidth in this article is evaluated ``ideally'' with considering the ac-side filter of inverter only but ignoring the influence of power networks, i.e., ideal current bandwidth.}

% \subsection{Summary of Case Studies}

% The eigenvalue and participation analysis based on descriptor state space properly coincides with the time-domain simulations based on EMT models. This validates both the feasibility and accuracy of using the descriptor state space to model multi-bus power systems with SGs and IBRs. In addition, it is worth highlighting that the derived state space model can not only predict the system stability and dynamics, but also clearly indicate the root cause of oscillation modes because all states are tracked during the modeling procedure, as discussed in \sectionref{Section:Superiority}.

% =========================
% Conclusions
% =========================

\section{Conclusions} \label{Section:Conclusions}

The descriptor state space method has been introduced for modelling multi-bus, multi-apparatus power systems composed of many subsystems. It compliments conventional state space analysis and can model both proper and improper systems. This feature solves the problem of port mismatches that can arise between subsystems with incompatible definitions of inputs and outputs. The states are always tracked during the modeling procedure so that a state of whole-system model can always be mapped to a physical state in one of the subsystem. With this preservation of states, participation analysis can be applied which enables analysis of root causes of instability and oscillatory  modes such that indications for parameter tuning in the physical plant can be provided. Algorithms for manipulation of descriptor state space subsystems are derived, including formation of model inverses, connections of subsystems, and other transformations. Numerical calculations and EMT simulations of example power systems were used to validate the theoretical advances.

% ========================================
% Appendix
% ========================================

\appendices

\section{Participation Analysis of Descriptor State Space} \label{Appendix:Participation}

The participation analysis of descriptor state space can be derived similarly to that of standard state space in Chapter 12 in \citeref{kundur1994power}, as briefly shown next. For a descriptor state space model in \equref{Equ:Dss}, we can get the generalized right and left eigenvector matrices $[\phi]$ and $[\psi]$ as \citeref{skogestad2005multivariable,matlabEig}
%\begin{equation}
%    \begin{aligned}
%    & [\phi] = \begin{bmatrix} \phi_1 & \phi_2 & ... & \phi_i & ...\end{bmatrix}
%    \\
%    & [\psi] = \begin{bmatrix} \psi_1 \\ \psi_2 \\ ... \\ \psi_i \\ ... \end{bmatrix}
%    \end{aligned}
%\end{equation}
\begin{equation}
    \begin{aligned}
    [\phi] = \begin{bmatrix} \phi_1 & \phi_2 & ... & \phi_i & ...\end{bmatrix}
    ,~
    [\psi] = \begin{bmatrix} \psi_1 \\ \psi_2 \\ ... \\ \psi_i \\ ... \end{bmatrix}
    \end{aligned}
\end{equation}
where $\phi_i$ and $\psi_i$ are the $i$th right and left eigenvectors. These two eigenvector matrices satisfy
%\begin{equation} \label{Equ:DssEigEqu}
%\begin{aligned}
%    & A [\phi] = E [\phi] [\lambda]
%    \\
%    & [\psi] A = [\lambda] [\psi] E 
%    \end{aligned}
%\end{equation}
\begin{equation} \label{Equ:DssEigEqu}
\begin{aligned}
    A [\phi] = E [\phi] [\lambda],~
    [\psi] A = [\lambda] [\psi] E 
    \end{aligned}
\end{equation}
where $[\lambda]$ is the eigenvalue matrix as
%\begin{equation}
%    [\lambda] = \begin{bmatrix} \lambda_1 &&&&& \\ & \lambda_2 &&& \\ && ... && \\ &&& \lambda_i & \\ &&&& .... \end{bmatrix}
%\end{equation}
\begin{equation}
    [\lambda] = \text{blkdiag}(\lambda_1,\lambda_2,...,\lambda_i,...)
\end{equation}
If only focusing the $i$th eigenvalue, we can re-write \equref{Equ:DssEigEqu} as
%\begin{equation} \label{Equ:EigVec}
%    \begin{aligned}
%    & A \phi_i = E \phi_i \lambda_i
%    \\
%    & \psi_i A = \lambda_i \psi_i E
%    \end{aligned}
%\end{equation}
\begin{equation} \label{Equ:EigVec}
    \begin{aligned}
    A \phi_i = E \phi_i \lambda_i,~
    \psi_i A = \lambda_i \psi_i E
    \end{aligned}
\end{equation}
Differentiating the first equation in \equref{Equ:EigVec} with respect to $a_{kj}$ (i.e., the element of $A$ in $k$th row and $j$th column) yields
\begin{equation} \label{Equ:DssPfEqu_}
    \frac{\partial{A}}{\partial{a_{kj}}} \phi_i 
    + 
    A \frac{\partial{\phi_i}}{\partial{a_{kj}}} 
    = 
    E \phi_i \frac{\partial{\lambda_i}}{\partial{a_{kj}}}
    +
    E \frac{\partial{\phi_i}}{\partial{a_{kj}}}\lambda_i
\end{equation}
Left-multiplying by $\psi_i$, and noting the second equation in \equref{Equ:EigVec}, we can re-write \equref{Equ:DssPfEqu_} as
\begin{equation}
    \underbrace{\psi_i E \phi_i}_{1~\text{or}~0} \frac{\partial{\lambda_i}}{\partial{a_{kj}}} 
    = 
    \underbrace{\psi_i \frac{\partial{A}}{\partial{a_{kj}}} \phi_i}_{\psi_{ik} \phi_{ji}}
\end{equation}
For normalized $[\phi]$ and $[\psi]$, the left-hand side equals to 1 (when $\lambda_i$ is a practical mode) or 0 (when $\lambda_i$ is a virtual mode). The right-hand side equals to $\psi_{ik} \phi_{ji}$, which coincides with the widely-known participation factor used in power engineering community \citeref{kundur1994power}.

% =========================================================
% End
% =========================================================

\ifCLASSOPTIONcaptionsoff
  \newpage
\fi

\bibliographystyle{IEEEtran}
\bibliography{Paper}

\end{document}